\documentclass[runningheads]{llncs}
\usepackage{graphicx}
\usepackage{booktabs}
\usepackage{tabularx}
\usepackage{hyperref}
\begin{document}
\title{Two case studies on implementing best practices for Software Process Improvement\thanks{This work is part of a project that has received funding from the European Union's Horizon 2020 research and innovation programme under Grant Agreement No. 856726 (GN4-3). \protect\\The scientific/academic work is financed from financial resources for science in the
years 2019-2022 granted for the realisation of the international project co-financed
by the Polish Ministry of Science and Higher Education.}}
\titlerunning{Two case studies on implementing best practices for SPI}
%

\author{Bartosz Walter\inst{5,1} 
\and
Branko Marović\inst{2} 
\and
Ivan Garnizov\inst{3}
\and
Marcin Wolski\inst{1}
\and
Andrijana Todosijevic\inst{4}
}
\authorrunning{B. Walter, B. Marović, I. Garnizov, M. Wolski, and A. Todosijevic}
%
\institute{PSNC, Poznań, Poland
\email{\{bartek.walter,marcin.wolski\}@man.poznan.pl}
\and
University of Belgrade, Belgrade, Serbia,
\email{branko.marovic@rcub.bg.ac.rs}
\and
Friedrich-Alexander-University, 
Erlangen, Germany,
\email{ivan.garnizov@fau.de}
\and
AMRES, Belgrade, Serbia,
\email{andrijana.todosijevic@amres.ac.rs}
\and
Poznań University of Technology, Poznań, Poland
}
\maketitle              
\begin{abstract}
Software Process Improvement requires significant effort related not only to the identification of relevant issues and providing an adequate response to them but also to the implementation and adoption of the changes. Best practices provide recommendations to software teams on how to address the identified objectives in practice, based on aggregated experience and knowledge. In the paper, we present the G\'{E}ANT experience and observations from the process of adopting the best practices and present the setting we have been using.

\end{abstract}
\section{Introduction}
\label{sec:intro}


The process of software construction, development and maintenance is probably more complex than many processes we observe in other areas of engineering. There are different reasons for that, starting from dynamically changing requirements, through compatibility issues, up to the evolving environment and the complexity of the supporting tools ecosystem. As a result, processes in the software domain are not fixed and need to be constantly updated, optimised and improved, to remain competitive in the global market and meet quality expectations. Regardless of the invested efforts, many software projects still exceed one or a few of the typical business constraints, related to cost, quality and time. It reveals a  gap between the expectations and reality, but also a large room available for improvement.

Software Process Improvement (SPI) embraces all efforts aimed at understanding the software process as it is being used within an organisation, and driving the implementation of changes to that process, intended to achieve specific goals, such as increasing development speed, achieving higher product quality or reducing costs. Usually, the business and technological constraints are mutually contradictory, so the objective is to find an acceptable balance between the expectations of stakeholders, constraints concerning cost and quality, and the capacity and capability of the development team. Until now, various approaches related to the design, implementation and evaluation of SPI have been identified, studied and verified~\cite{Unterkalmsteiner2014}. A typical SPI process has many stages, starting from the identification of objectives, via designing changes, operational procedures and metrics for monitoring, up to implementing them in practice~\cite{Aysolmaz2011}. Usually, SPI endeavours are not linear, but iterative~\cite{Stojanov2016}, resulting in a highly reactive and continually improving process driven by the collected feedback. It will not only help in improving the process, but can reorient the organisation to continuously seek opportunities for further optimisation.

In previous papers, we have presented various aspects of the SPI-related challenges, approaches and results of developing and operating an SPI process in G\'{E}ANT, a distributed multi-national European project focused on innovation and exploration of new research directions within networking and software~\cite{Stanisavljevic2018,Walter2020euro,eurospi2021}. It faces specific challenges, which needs to be reflected in the SPI process. As a result, it is based on two pillars: a maturity model (SMM) which defines objectives that need to be made, and a catalogue of best practices (CBP) that are aligned with the SMM and recommend methods of addressing the objectives. Best practices appear to be a promising way of promoting proven and well-aligned approaches, techniques and tools.

In this work, we report two examples of implementing best practices in different software teams in G\'{E}ANT. The early results show that teams can properly recognise their needs,  address them using the framework, and monitor the implementation with the progress metrics. 

The remainder of this work is structured as follows. In Sec.~\ref{sec:relwork} we make a brief overview of the literature concerning the SPI, its implementation and support from the best practices.
In Sec.~\ref{sec:background} we briefly provide information about G\'{E}ANT, its structure and constraints concerning software teams. We also present the best practices framework designed and developed to manage the SPI processes.
Two other sections,~\ref{sec:projectA} and \ref{sec:projectB} show how it has been implemented in two different projects, with preliminary results. Observations from that implementation are available in Sec.~\ref{sec:discussion}, and in Sec.~\ref{sec:summary} we summarise the work and set the directions for the next steps.

\section{Related work}
\label{sec:relwork}


Software processes need to be continuously evaluated and optimised in order to better fulfil the expectations of all the stakeholders of software projects ~\cite{fuggetta2014software}. The maturity of software teams is an established concept used for reflecting their ability to effectively address and implement all requirements and fulfil objectives in their tasks. Maturity models (MMs) identify the objectives relevant for specific areas and evaluate the performance with respect to these objectives. Garc{\'{i}}a-Mireles, Moraga and Garc{\'{i}}~\cite{Garcia-Mireles2012} reported more than 50 software process capability/maturity models (SPCMM), 35 different maturity models related to the discipline of information systems, and more than 150 models that assess the maturity of IT service capability, strategic alignment of innovation management, program management, knowledge management and enterprise architecture. 

Effective methods of addressing the objectives and goals defined in maturity models require also adequate guidance and recommendations in the form of best practices~\cite{swebok}. A lot of patterns for the successful adoption of SPI practices have been recognised so far ~\cite{singer2013improving}. Best practices for the successful design and implementation of lightweight software process assessment methods are investigated by Zarour et al. ~\cite{Zarour2015}. In some of the models, a large part of the SPI success is achieved through the adoption of best practices~\cite{Bayona2019}.

In~\cite{WalterMGWT21} we explained the approach to monitoring the adoption of best practices accompanied by preliminary observations of the process improvements.
\section{Background}
\label{sec:background}


G\'{E}ANT is a large pan-European project that initially originated from networking, but now it has become a leader in delivering innovative solutions for research and academia, by embracing complex software components and systems and defining services on top of the network. As a  software organisation, it features an innovation-targeting culture combined with a specific, custom approach to software development. Software products and components are developed in small independent teams (SDTs) that are internally diversified by the nationality of members, native culture of their organisations, remote distance etc. In addition, the teams are largely self-organising, with a large extent of freedom in defining specific processes, or adapting them to their needs. Such high diversity creates a friendly ground for innovations, but also poses a risk for business-driven constraints, effective coordination of work and managing the outcomes~\cite{Stanisavljevic2018}. 

The specifics in the objectives and the federated organisation in the G\'{E}ANT project have a significant impact on the working environment and coordination of effort in the software teams.
They adhere to a Product Life-cycle Management (PLM) framework and use common tools, but they can internally define their own processes, techniques and settings~\cite{wolski_rodwell}, which is a significant challenge from the process-oriented point of view.

\subsection{Common Best Practices}

A large part of SPI in G\'{E}ANT is founded on Common Best Practices (CBPs)~\cite{eurospi2020,eurospi2021}. Unlike the goals identified in the Software Maturity Model (SMM)~\cite{eurospi2019}, which set objectives for the teams, but do not indicate the ways of addressing them, the CBPs also provide directional guidance on \emph{what} could be done in order to achieve the specific goals defined in the SMM. The guidance is based on the experience of other teams and the activities that have been shown to succeed in G\'{E}ANT's setting.

Currently, 24 practices have been identified, described and summarised in a catalogue of best practices~\cite{eurospi2021}. They cover all relevant aspects of software development in G\'{E}ANT that have been recognised during the analysis of needs and objectives~\cite{Wolski2017}, and additionally include areas related to the team organisation and product maintenance. For brevity, in Table~\ref{tab:cbp} we present only the ones that have been adopted by the teams presented and discussed in Sec.~\ref{sec:studies}. 

Apart from other simple attributes, each practice is described by several recommendations, activities, and metrics. 

\textbf{Recommendations} --
Each CBP is detailed into 2-7 recommendations that cover different aspects of the best practice. The recommendations prescribe objectives of a best practice process along with recommendations for tools, the type of tools, the skills and/or procedures that support the process. They point to the key aspects of the best practice. 

\textbf{Activities} --
Each recommendation may be associated with a number of activities. An activity is an actionable statement that contributes to the fulfilment of the associated recommendation. It is generic enough to be relevant for many SDTs, regardless of their size and experience. Activities are not steps to be executed in a sequence (e.g., they may be done in parallel), but individual pieces of work that need to be performed in order to fully address the recommendation. Of course, the practical implementations may vary between development teams, as the best practices leave room for their operationalisation, adaptation and customisation.

\textbf{Metrics and their values} --
Metrics address individual, low-level concepts that can be directly measured or marked as achieved or unsettled. Metrics reflect the current status of a single aspect of implementing the practice; not just for the development team, but also for SwM. They are substantial and trustworthy, so that management and external reviewers can rely on the evaluation of these metrics when concluding about the achievements, maturity and/or readiness of the SDT in specific topics or with regard to the specific risks related to the activities reviewed in the respective best practice.

\subsection{Structure of the SPI project}
The software governance and management team (SwM), formed as a part of one of G\'{E}ANT work packages, is the entity that coordinates the process of defining, managing, implementing and maintaining the best practices framework and the catalogue. SwM and its experts are also responsible for engaging the teams promoting SPI within G\'{E}ANT and collecting feedback, which is further used for improvement. 

Software development teams (SDTs) voluntarily participate in the implementation of best practices, however, this is positively acknowledged by the management and appreciated as one of the recommended self-improvement methods for all software teams.

\begin{table}[]
\caption{An excerpt for the Catalogue of Common Best Practices used in G\'{E}ANT. Presented metrics have been chosen for implementation in Projects A and B (see Sec.~\ref{sec:studies})}
\begin{tabularx}{\linewidth}{p{0.1\linewidth} p{0.9\linewidth}}
\toprule
ID      & Practice                                                                    \\ 
\midrule
BP-A.1  & Identify an initial group of stakeholders and iteratively refine it \\
BP-A.2  & Elaborate communication strategy for stakeholders                 \\
\midrule
BP-C.2  & Identify relevant quality characteristics and test conditions and provide verification criteria for them \\
BP-C.3  & Elaborate and maintain a quality plan for the project               \\
\midrule
BP-E.4  & Define a procedure for deploying changes to running services \\ 
\bottomrule
\end{tabularx}
\label{tab:cbp}
\end{table}



\subsection{Product Life-Cycle Management and Best Practices}
Although software development teams are given the freedom to choose on their own the methodologies for the development and organisation of the effort, certain boundaries are set to preserve the business perspective and high-level goals of the G\'{E}ANT project. These define the Product Lifecycle Management process imposed and supervised by the Product Management Team. The process is split into several phases, for which the development teams are required to demonstrate a certain level of readiness with the product, maturity in delivery, documentation and engagement from the user community in order to be able to gain further support for their work in the next phase. Here the benefit of Best Practices was well recognised, as they also assert specific levels of maturity in the process of software development and deal with various related risks, which often match various concerns and risks identified on the project management side.

\section{Case studies}
\label{sec:studies}

Based on the previously described work, we started implementing the best practices as a method for improving software processes in G\'{E}ANT. Below we briefly present two projects, along the the description of the undertaken activities.

\subsection{Methodology}
In both cases we applied a similar,  repetitive approach that followed a number steps in each software team:

\begin{enumerate}
\item Identify the needs of the subject team
\item Identify and scope the area of improvement in the context of the G\'{E}ANT CBPs.
\item Agree with the team on the approach, by addressing its context and by including the specific actions and metrics used for monitoring.
\item Monitor the progress and refine the approach (if needed).
\item Conclude the process and evaluate results.
\end{enumerate}

For monitoring the CBP adoption we have developed a dedicated feature in the G\'{E}ANT Software Catalogue~\cite{eurospi2021}\footnote{\url{https://sc.geant.org}}. It allows the assessors from SwM for collecting and reporting individual assessments, based on metrics defined for best practices, and tracking their changes.



Each evaluation is described by several attributes, such as a timestamp or a scale. Values on the scale show how well a given team has adopted the practices. The values can also be supplemented by a comment, which gives a more thorough picture of the process at the moment. Having the evaluation recorded, a team member can get an insight into practices implemented by the team or see the progress in practices implementation within a specific period.

\subsection{Project A}
\label{sec:projectA}



\subsubsection{Overview of the project}  -- 
The project is a software-based service platform for assessment and monitoring of the WiFi performance and user experience in campus areas. End-user devices exchange short anonymised data packets to collect multiple network performance statistics. The architecture comprises of a server side for measurement, data collection and GUI with monitoring modules. 
The team provides the software as a product, but also a number of services: support and a live demo environment.

\subsubsection{Identified needs and expectations}  -- 
During an opening meeting, the team indicated a number of potential areas for  improvement, referring to diverse areas, from requirements to software deployment. After discussion, the problem of managing software configuration and building it for different customers was found to be essential. The team struggled with delivering updates to many releases of the product, which resulted in mistakes and delayed delivery. 

\subsubsection{Best practice chosen for implementation} -- 
After the review, a BP-E.4 best practice has been identified to match the team's needs. It is focused on the process of implementing changes to services in operation, which requires a good understanding of the explicit and implicit dependencies of the running system, the risks associated with the management of operational data and the tight coordination between the different operational teams that provide supporting generic services for example, but not limited to, network connectivity or operating system management in an attempt to minimise the possible running service downtime.  

The best practice specifies three recommendations:
\begin{enumerate}
    \item Identify services that could be affected by the subject change;
    \item Define a deployment procedure that minimises the impact on other services;
    \begin{enumerate}
        \item  Consult the procedure with owners of affected services;
        \item Prepare and verify rollback routines (at least for the core parts of the deployment procedure) that allow for reversing the deployment in a safe way;
        \item Define the deployment schedule in collaboration with the owners of affected services.
    \end{enumerate}
    \item Make a simulated deployment for the change, if necessary.
    \begin{enumerate}
        \item Involve the owners of the affected services;
        \item Monitor the simulation and make notes concerning the observed deviations from the procedure;
        \item Update the procedure, if needed.
    \end{enumerate}
\end{enumerate}

\subsubsection{Results}  -- 
Following the established SPI process, the SwM team conducted an initial assessment of the process that could serve as a basis for monitoring.

The SDT developed an internal service to track the product versions ran by customers to identify those who are actively using it. In the process, they also clarified the internal release management process, which enabled them to perform and collect the version checks (automated or manual) from the active instances.

As follows from the second assessment, the team did improve their process and is now running an ordered, disciplined and data-based process related to the version management, properly, tracking the product use and associating versions with specific customers and their issues, needs and questions (see Table~\ref{tab:projectAmetrics}).

\begin{table}[]
\label{tab:projectAmetrics}
\caption{Results of evaluation for the BP-E.4 best practice in Project A}
\centering
\begin{tabular}{lrrr}
\toprule
Metric  & \multicolumn{3}{c}{Assessments} \\
        & M1         & M5         & M15\\ 
\midrule
General evaluation & 3/7 & 6/7 & 7/7\\
\midrule
Orderly collection of new feature requests
& YES            & YES               & YES              \\
Procedure for choosing features to be added to the new release
& YES             & YES              & YES              \\
Predefined release schedule
& YES             & YES              & YES              \\
\midrule
Communication channel for informing users about updates                                                            & NO             & YES              & YES              \\
Mechanism for performing or nudging updates                                                            & NO             & YES              & YES              \\
Tool or method for collecting the data about used versions and updates                                                            & NO             & YES              & YES              \\
Collected data is used in analysis and planning                                                            & NO             & NO              & YES              \\
\midrule
\bottomrule
\end{tabular}
\end{table}

The changes resulted from additions to delivery, deployment, monitoring, user support, and release planning that were articulated during the joint sessions of SwM and Project A teams. These additions were expressed as clear and straightforward software features, utilities and procedural changes, which enabled the developers and supporters to adopt them as practical and valuable modifications, easily implement the needed elements and get immediate improvements. All devised and agreed interventions have been implemented; the data is collected as a part of the monitoring process and periodically analysed. 

An internal procedure for deploying, validating or reverting the changes was established and applied. Information about the new releases and summaries of contained changes are timely distributed to registered users. Furthermore, the project team has established the process of collecting records that describe and track features and issues reported by users and testers. The instances of the software update automatically or inform the user about available updates and assist in their download and installation. The team also developed a method for collecting the data about used the usage and versions of WiFiMon instances and the rollout of updates. The tracked parameters are the numbers of active instances and downloads and the percentage of users running the latest version. Data about changes, their availability, deployments, active instances and related user behaviours and response times are used in analysis and planning to enhance release management, deployment process or communication with users. If new problems or opportunities for improvement are identified, they will be used to further refine the team's procedures, software tools and metrics, and also the related best practices.

\subsection{Project B}
\label{sec:projectB}

\subsubsection{Overview of the project}  -- 
Project B is an internal online catalogue that includes comprehensive information about software projects developed and maintained in G\'{E}ANT. Unlike other similar systems, it does not rely on data entered manually by operators or users, but rather actively looks for changes in the available data sources (such as software development tools) and updates its internal databases. The offered data is diverse; for example, it displays the description and current status of the project, its current and previous staff, external organisations involved in it, but also development-related activity in the code repository, quality-related reports, the employed technology stack, and also some specific SPI-related information. In addition, the project is also endowed with a full-text search engine, which helps the users in finding the relevant pieces of information. From the functional point of view, Project B is a central informational hub with comprehensive data about the projects.

\subsubsection{Identified needs and expectations}  -- 
Project B is managed from the beginning by a consistent, stable team of developers, with a clear vision of the product. As a result, several processes and software engineering practices and properly implemented, addressed and aligned. However, the team decided to participate in the best practice assessment with an objective to identify deviations, quality hotspots and opportunities for improvement.

Specifically, the team decided to focus on two specific goals: the management of stakeholders, and the elements of quality assurance. Both of them have been partially implemented by achieving and documenting some objectives, while some aspects have not been addressed yet. The implementation of best practices provided a good opportunity for improving the status vis-à-vis these goals.

\subsubsection{Best practice chosen for implementation}  -- 
A number of meetings and interviews allowed for the identification of the practices to be assessed and implemented in the beginning: 
BP-A.1: Identify an initial group of stakeholders and iteratively refine it.

The BP-A.1 practice is focused on having a closely engaged group of stakeholders, who would and could help in collecting requirements and user needs. The stakeholders could also contribute to other areas of the project, e.g., identification of risks (BP-C.1) or definition of user acceptance tests (BP-C.4).

Specifically, there are two main recommendations formulated for that practice:
\begin{enumerate}
    \item Identify an initial group of stakeholders
    \begin{enumerate}
        \item Consider teams, organisations or individuals that could be affected or could impact the project.
        \item Look for similarities to other projects, either previous or current.
        \item Look for a dominant stakeholder, who is mostly interested in the outcome of the project.
    \end{enumerate}
    \item Maintain (update) the group of stakeholders
    \begin{enumerate}
        \item     Publish the list of stakeholders and their representatives.
        \item Periodically update (involve and retire) the group of stakeholders.
        \item Apply snowballing to identify new stakeholders.
        \item Categorise the stakeholders with respect to their relevance for the project
        \item Identify possible relationships between stakeholders
    \end{enumerate}
\end{enumerate}

\noindent In Project B, this aspect has been addressed by creating and maintaining a list of stakeholders. The list is managed by the team leader and periodically reviewed/updated. Each communication targeted to the stakeholders is also recorded to monitor the frequency of meetings as well.
The part of the practice that was not implemented concerns the prioritisation of stakeholders, and all of them were considered equally involved in the project. The recommended solution would be to identify a group of key stakeholders to be consulted regularly, and the remaining ones, who would be only informed about the changes and the progress. 



\subsubsection{Results}  -- 
Implementation of those best practices has been monitored with a set of metrics that were adapted  from the predefined ones for that practice. The results of two evaluations are reported in Table~\ref{tab:projectBmetricsA1}. They are rather indicative than conclusive, and provide some information on how the assessment is being made.

\begin{table}[]
\caption{Results of evaluation for the BP-A.1 best practice in Project B}
\centering
\begin{tabular}{lrr}
\toprule
Metric  & \multicolumn{2}{c}{Assessments} \\
        & M1         & M2         \\ 
\midrule
General evaluation & 4/5 & 5/5 \\
\midrule
Are stakeholders prioritised?                                                            & NO             & YES               \\
Is the list of stakeholders updated on a regular basis?                                 & YES            & YES               \\
Are the stakeholders' needs tracked to requirements?                                    & YES             & YES               \\ 
\midrule
Number of identified and contacted stakeholders & 14  & 14  \\
Time since contact with a stakeholder (about requirements) {[}days{]} 
    & N/A               & N/A               \\ 
Time since contact with a stakeholder (about UAT) {[}days{]}          
    & N/A               & N/A               \\ 
\bottomrule
\end{tabular}
\label{tab:projectBmetricsA1}
\end{table}

There are three simple Boolean metrics. They directly refer to facts, not evaluation, which makes them more objective. However, they also carry limited information, which can be supplemented with a description. In that case, the assessor provided no remarks.
Additionally, there are also three regular metrics. They capture the number of stakeholders that have been identified by the team, and the time since contacting any stakeholder with respect to requirements and UATs, which reflects the activity in communication with them. The two time-related metrics are not being monitored, as the data has not been collected so far. The team also decided that they are not relevant in that case.

During the first assessment it has been found that a relatively large group of stakeholders is managed in the same way, which did not properly reflect the real needs. The SwM and Project B teams have decided to refine the list of stakeholders to identify key stakeholders and involve them more closely in the project. As a result, a dedicated list of key stakeholders was introduced, which was appreciated during the second assessment. 


Currently, all recommendations for the practice have been implemented, and now it needs to be only monitored. If new opportunities for improvement would be identified, then they could result in defining another, more refined best practice.

\section{Concluding observations}
\label{sec:discussion}

Two projects are not sufficient to make conclusions concerning the effects of applying best practices as drivers for SPI improvement. However, we can summarise the current status of the project with some concluding observations.

Each practice has been linked with an initial set of predefined metrics that could be used for monitoring. The metrics are diverse with respect to various dimensions addressed by the practice and provide a starting point for measuring progress. However, the set is only a proposal and is not closed: it can be easily extended by defining new metrics or modifying existing ones, depending on the context, needs and capabilities of individual development teams. Each SPI implementation project is initiated at a meeting, during which both the SDT and SwM teams agree on how it would be monitored, which also includes the selection of specific metrics. It ensures that only relevant parameters which reflect the specifics of the project are measured.

The metrics are a relatively new element of the framework. Unlike directional recommendations included in each best practice, metrics are specific and set a kind of a roadmap that facilitates its implementation. They reflect different dimensions of the corresponding SMM goal and the associated CBP, so they help both the development team and the assessors in monitoring the progress. However, metrics should not be considered as a tool for evaluating the team, but rather support in monitoring the implementation of best practices.

Introducing metrics to the framework is at least partially successful. On one hand, the measurements using the metrics are still subjective and should be conducted by trained support staff, and in close collaboration with the software development team. On the other hand, even in their current form they provide real help to the assessors and let them focus on specific objectives, goals and process dimensions defined by a given best practice. Therefore, they indicate the ways of interpreting the practices in the desired direction for implementation, largely reducing ambiguities in their interpretation. 

Due to subjective factors in the metrics and necessary per-case modifications, the best practices framework cannot be used to compare the maturity of different projects and their progress in implementing them. Cross-project studies would be inherently biased by different metrics for the evaluation and differences in teams' contexts, e.g., the team structure and size, adopted methods of communication, external needs, the tools used, etc. However, the metrics can still be used for monitoring projects and conducting longitudinal studies.

\section{Summary}
\label{sec:summary}

In this work, we have presented two examples for implementing the SPI project based on the best practices framework, from the identification of needs and expectations to monitoring. In particular, the monitoring process was founded on metrics, which deliver objective and quantitative information about the progress. While the implementation process in other teams is still not finished, we were able to validate the entire process with real software development teams and collect feedback.

First, the two projects delivered several insights concerning ways of engaging teams in the SPI effort, the methods of extracting their real needs with respect to process maturity, and provided hints about planning the implementation process. Based on that, future implementations could be more streamlined and less effort-demanding.

Overall, the structure of the SPI project appeared effective. In each case, the work coordination is in the hands of the SwM team, which is also responsible for planning the work and improving the SPI framework.
In principle, collaboration with each of the software teams is managed by one lead expert, supported by another person. This fosters fluent and effective communication within the SwM team, but also provides a smooth take-over in case of unavailability of the lead expert.

Thanks to the specific recommendations and related metrics, teams have become more aware of the complexity of SPI activities and their multi-dimensional impact on other processes. They also understood the relationship between intuitive, but quite generic recommendations and the specific metrics that reflect the progress. In many cases, the teams have considered the metrics not as progress-tracking tools, but rather as an illustration of how the framework worked. 

The implementation has become a process on its own, with inputs, phases, expected and actual outcomes. That also helped both the development teams and the SwM team to stick on track and complete the implementation of the best practice successfully.

Future work includes the wider adoption of best practices, with an objective of collecting more data about the process performance, so that the conclusions could be founded on more stable empirical evidence.

\bibliographystyle{unsrt}
\bibliography{gsmm}

\end{document}